\documentclass[apj]{emulateapj}
\usepackage[colorlinks = true, linkcolor = blue, urlcolor  = blue, citecolor = blue, anchorcolor = blue]{hyperref}
\usepackage{hyperref}
\usepackage{multirow}
\usepackage{amsmath}
\usepackage{float}
\usepackage{multirow}

\usepackage{comment}

\shorttitle{Systematic variations of macrospicule properties observed by SDO/AIA over half a decade}
\shortauthors{T. S. Kiss et al., 2016}

\begin{document}

\title{Systematic variations of macrospicule properties observed by SDO/AIA over half a decade}

\author{T. S. Kiss \altaffilmark{1,2} , N. Gyenge\altaffilmark{1,3}, 
 R. Erd\'elyi\altaffilmark{1*}}
\thanks{\altaffilmark{*}e-mail: robertus@sheffield.ac.uk}
\affil{\altaffilmark{1}Solar Physics and Space Plasmas Research Centre (SP2RC), School of Mathematics and Statistics, University of Sheffield\\
Hicks Building, Hounsfield Road, Sheffield, S3 7RH, United Kingdom\\
\altaffilmark{2}Department of Physics, University of Debrecen,
Egyetem t\'er 1, Debrecen, H-4010, Hungary\\
\altaffilmark{3}Debrecen Heliophysical Observatory (DHO), Research Centre for Astronomy and Earth Sciences, Hungarian Academy of Sciences\\
Debrecen, P.O.Box 30, H-4010, Hungary\\}

\begin{abstract}
Macrospicules (MS) are localised small-scale jet-like phenomena in the solar atmosphere, which have the potential to transport considerable amount of momentum and energy from the lower solar atmospheric regions to the Transition Region and the low corona. A detailed statistical analysis of their temporal behaviour and spatial properties is carried out in this work. By means of state-of-the-art spatial and temporal resolution observations, yielded by the Atmospheric Imaging Assembly (AIA) of Solar Dynamics Observatory (SDO), we constructed a database covering a 5.5-year long period, containing 301 macrospicules that occurred between June 2010 and December 2015 detected at 30.4 nm wavelength. Here, we report the long-term variation of the height, length, average speed and width of MS in Coronal Holes and Quiet Sun areas both in the northern and southern hemisphere of the Sun. This new database helps to refine our knowledge about the physical properties of MS. Cross-correlation of these properties show a relatively strong correlation, but not always a dominant one. However, a more detailed analysis indicates a wave-like signature in the behaviour of MS properties in time. The period of these long-term oscillatory behaviours are just under two years. Also, in terms of solar north/south hemispheres, a strong asymmetry was found in the spatial distribution of MS properties, which may be accounted for the solar dynamo. This latter feature may then indicate a strong and rather intrinsic link between global internal and local atmospheric phenomena in the Sun. 
\end{abstract}

\section{Introduction}
Our knowledge of the different layers and structures of the solar atmosphere has improved greatly in the last decades. However, the structural fine-scale details of the atmosphere still leaves many open questions (see the review papers of e.g. \citealt{judge2006, lipartito2014}). A challenging task is to identify and catalogue the small-scale, localised observable phenomena (e.g. bushes, fibrils, flocculi, grains, mottles, spicules, etc.) present in the chromospheric zoo in regards of their numbers and variety. These, often rapidly appearing and disappearing, fine structures are popularly observed in e.g. the He II ($\approx 30.4$ nm) emission line, H$\alpha$ ($\approx 656.28$ nm) and Ca II ($\approx 393.366$ nm) absorption lines.

One class of the set of highly dynamic phenomena are the spicules. They are abundant, spiky-like gas jets at the chromospheric solar limb \citep{secchi1877}. Their upward mass flux is about 100 times that of the solar wind \citep{sterling2000, depontieu2004, sterling2010}. Spicules are detectable both on the disc (often called as mottles) and at the limb at any given moment of time. Depending on their size and lifetime, spicules can be classed into two groups. Spicules with $7000 - 11000$ km length, 5 -- 15 min lifetime and 25 km s$^{-1}$ propagating speed are called "classic" or lately debatably referred to as type-I spicule \citep{beckers1968, zaqarashvili2009}. The other group is claimed to  contain the smaller and faster ($5000$ km average height, $50 - 100$ km s$^{-1}$ propagation speed) so-called type-II spicule (on disc often labelled as Rapid Blueshifted Excursiors -- RBEs referring to one of their distinct observable properties) with shorter lifetime, 10 -- 150 s \citep{depontieu2007, depontieu2012, sekse2012, kuridze2015}. 
Spicules, irrespective of their lifetime and speed, could play an important role in the energy and momentum transportation from the photosphere to the chromosphere or low corona. Their capability of guiding MHD waves are discussed by e.g. \cite{zaqarashvili2009}.

There are, however, much larger spicule-like, also abundant, dynamic jets detected to be present in the solar atmosphere: macrospicules (MS). 
The investigation of macrospicules reaches back at least 40 years. The first modern observation of macrospicules was carried out by \cite{bohlin1975}. With the 30.4 nm spectroheliograph onboard the SKYLAB mission, \cite{bohlin1975} identified 25 MS with length of 8" to 25" (some extreme cases were even with 30"-60"). Their lifetime is about 8 to 45 minutes without any reported correlation at that time between the length and lifetime. Most MS are formed inside coronal hole regions. Their name, initially, was suggested to be "EUV macrospiclues" after the wavelength band they were detected in.

\begin{figure*}[t]
	\centering
	\includegraphics[scale=0.23]{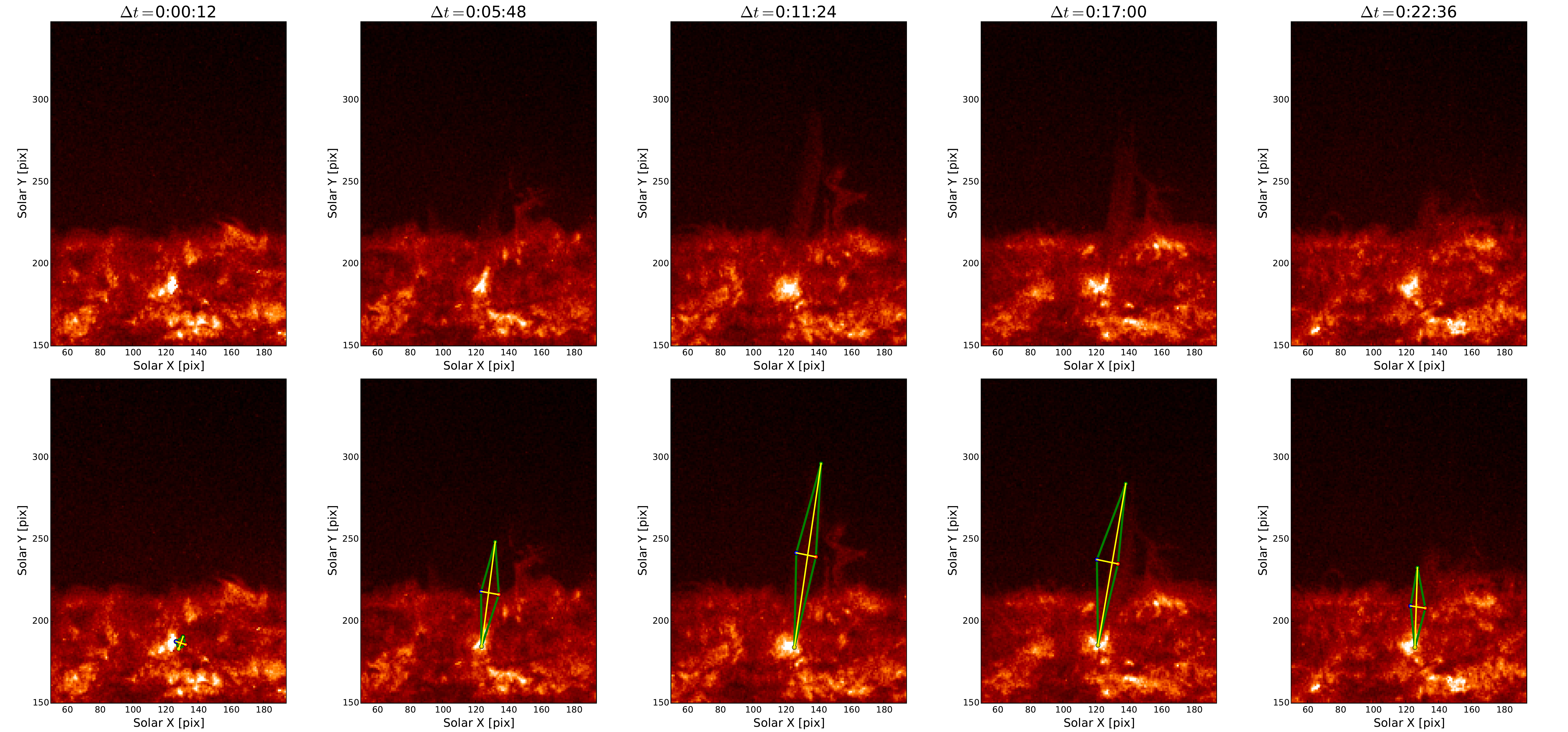}
	\caption{Series of images about an Quiet Sun macrospicule on SDO/AIA 30.4 nm images. The phenomenon occurred between 12:20 and 12:43 on 03.07.2012 on the western solar limb. The bottom panel images are overplotted by the tetragon assumption of MS. In the first column, the brightening is clearly visible, which may be the precursor of the jet itself.}
\end{figure*}

\cite{labonte1979} observed 32 macrospicules with the 25-cm aperture vacuum telescope at Big Bear Solar Observatory using H$\alpha$ and $\text{D}_3$ filters during an observation campaign of 122 hours. The length (8-33") and lifetime (4-24 min) of these MS were similar to those of reported in \cite{bohlin1975}. 
\cite{dere1989} investigated 10 MS by the UV spectroheliograph onboard the Spacelab-2 mission. The properties of the MS found were similar, again, to those reported in the previous two studies, but the main innovation is the development of temporal resolution: Spacelab-2 was able to observe macrospicules with a lifetime down to three minutes.

The breakthrough came with the age of high-resolution spacecrafts: Solar Heliospheric Observatory (SOHO, \citealt{fleck2000}) opened new avenues in solar physics with much greater temporal and spatial resolution as before. 
In the context relevant for our paper, we recall that \cite{pike1997} used SOHO/Coronal Diagnostic Spectrometer (CDS) to investigate the physics of MS with the first multi-wavelength observation and described them as multi-thermal structures. The authors claimed that X-ray jets and EUV MS are manifestation of the same physical phenomena underlying, which could be the source of the fast solar wind as well. They also described, first, an associated "pre-existing bright point", which can be a sub-flare brightening. 

In a follow-up study \cite{pike1998} employed SOHO/CDS to detect the rotational properties of MS. Blue and red-shifted emission was observed on either side of the axis of MS above the limb, which suggests the rotation of MS. The rotation velocities increase with height, so the macrospicules they observed were labelled "solar tornadoes". The separation of blue- and red-shifted regions on two sides of the macrospicule is clearly visible.
 
\cite{parenti2002} investigated the density and temperature of MS. Three datasets, taken by SOHO/CDS and SOHO/Normal Incidence Spectrometer (NIS) were constructed. The first dateset contains information about the MS, while the two others are averaged and used as a "background" to reduce the noise of the first one. The subtraced spectra showed a number of new properties: MS density was about $10^{-10} \text{ cm}^{-3}$ while the temperature is $ 3-4 \times 10^5$ K. The employed noise-reduction technique gave an opportunity to investigate the motion and the trajectory of MS: a maximum velocity was found to be about 80 km $\text{s}^{-1}$ during the entire lifetime. The maximum height, reached by MS, was reported to be $6 \times 10^4$ km above the solar limb. The average speed of falling back was about 26 km$\text{ s}^{-1}$. 

\cite{scullion2010} examined both on-limb and off-limb MS with the high-resolution spectroscopic capability of SOHO/Solar Ultraviolet Measurements from Emitted Radiation (SUMER) instrument. In case of two off-limb events, fast upward propagation was measured between mid-transition region (N VII - 765\AA) and the lower corona (Ne VIII - 770\AA) with $\approx 145 \text{ km s}^{-1}$. On-limb observations suggest that spicules can be precursors for macrospicules.

\cite{madrajska2011} analysed three macrospicules with four instruments (Hinode/SOT, EIS, XRT and SOHO/SUMER). These co-aligned images revealed that macrospicules do not seem to appear in spectral line formed over 300000 K.

On the theoretical side, \cite{murawski2011} carried out one of the first numerical simulations for MS as a velocity-shock in the Transition Region. They employed the FLASH code \citep{lee2009} to solve the two-dimensional MHD equations by implementing a VAL IIIC solar temperature model \citep{vernazza1981}. Many properties of simulated MS match the above mentioned, observed lifetime, length, velocity.

Another new era of MS observation has began with the launch of Solar Dynamics Observatory (SDO) in 2010 \citep{pesnell2012}. Onboard the spacecraft, the Atmospheric Imaging Assembly (SDO/AIA) generates terabytes of full-disc data with 0.6" spatial and 12 s temporal resolution at e.g. 30.4 nm wavelength \citep{lemen2012}.
\begin{figure*}[t]
	\centering
	\includegraphics[scale=0.2]{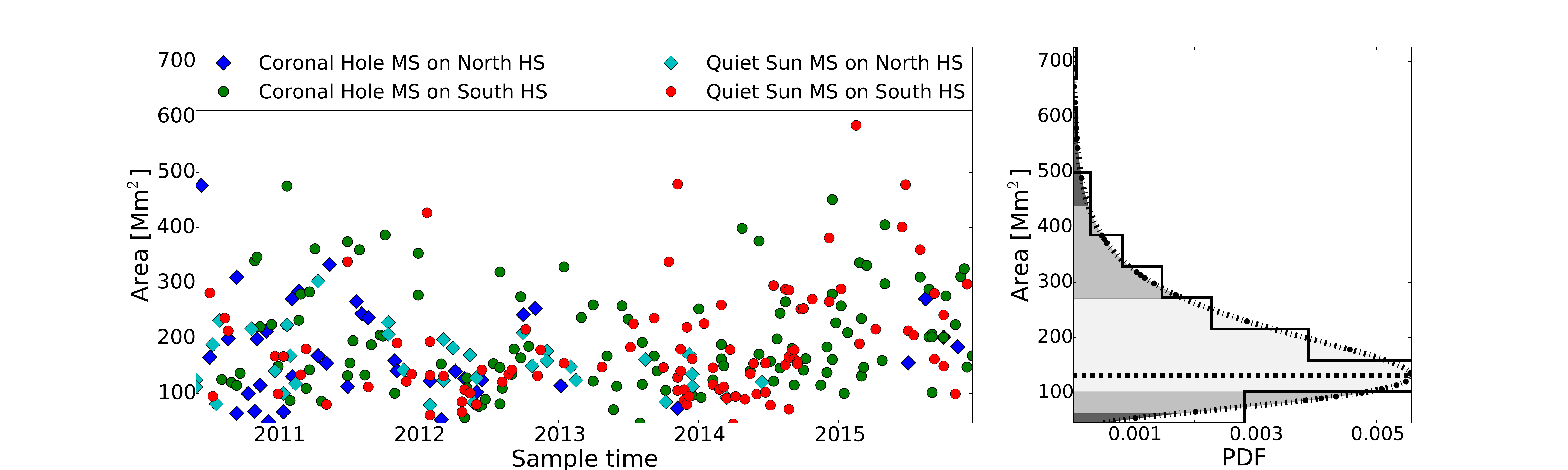}
	\includegraphics[scale=0.2]{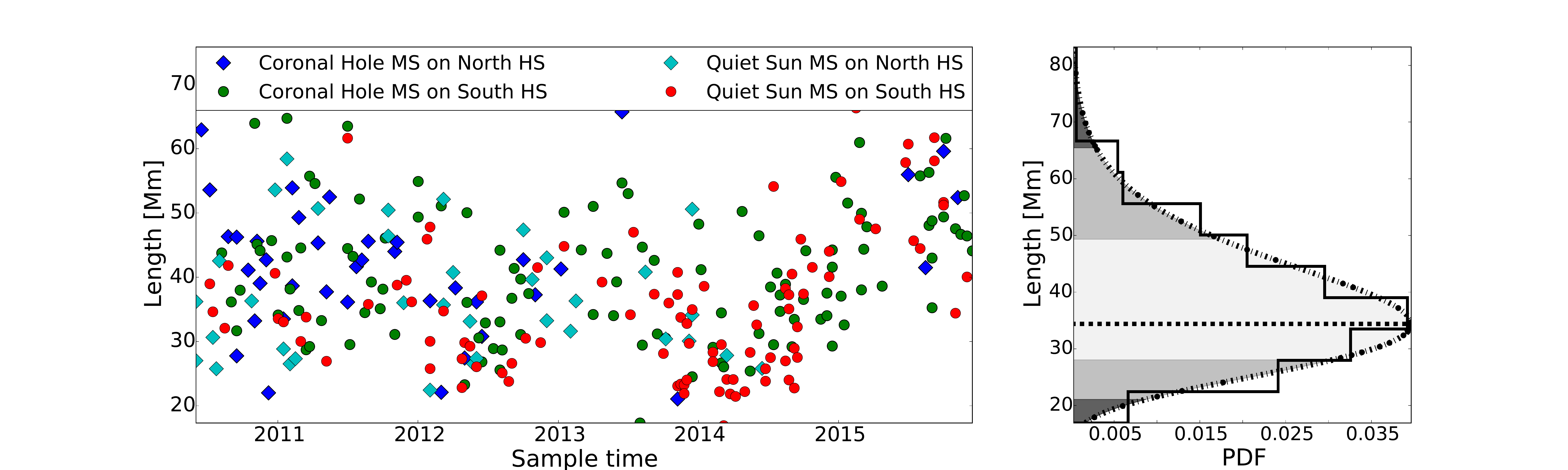}
	\includegraphics[scale=0.2]{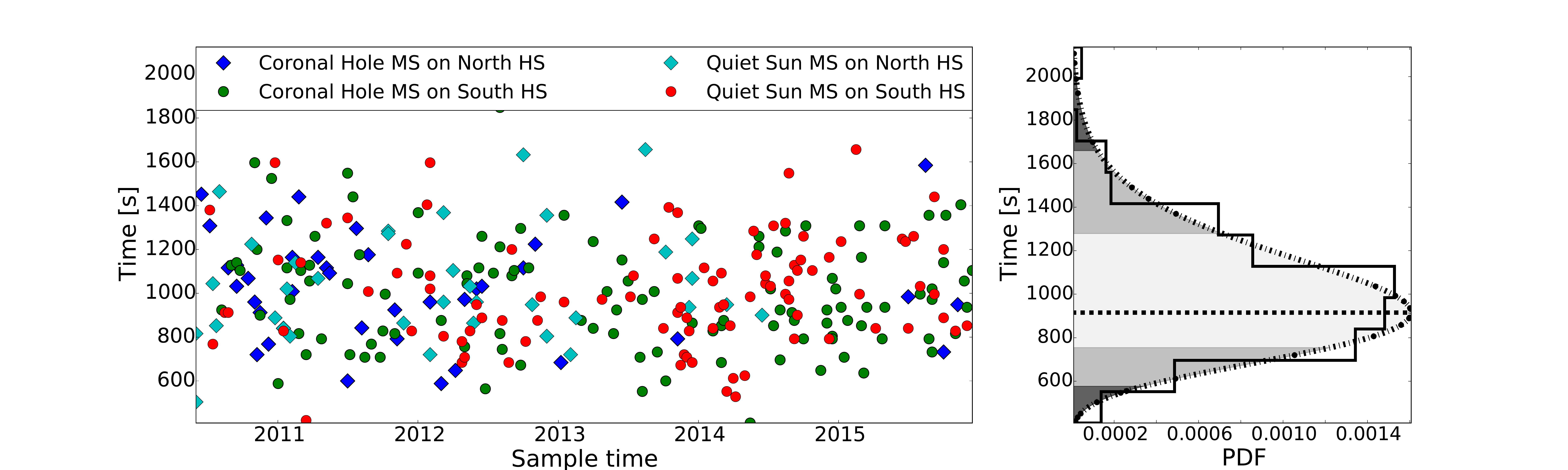}
	\includegraphics[scale=0.2]{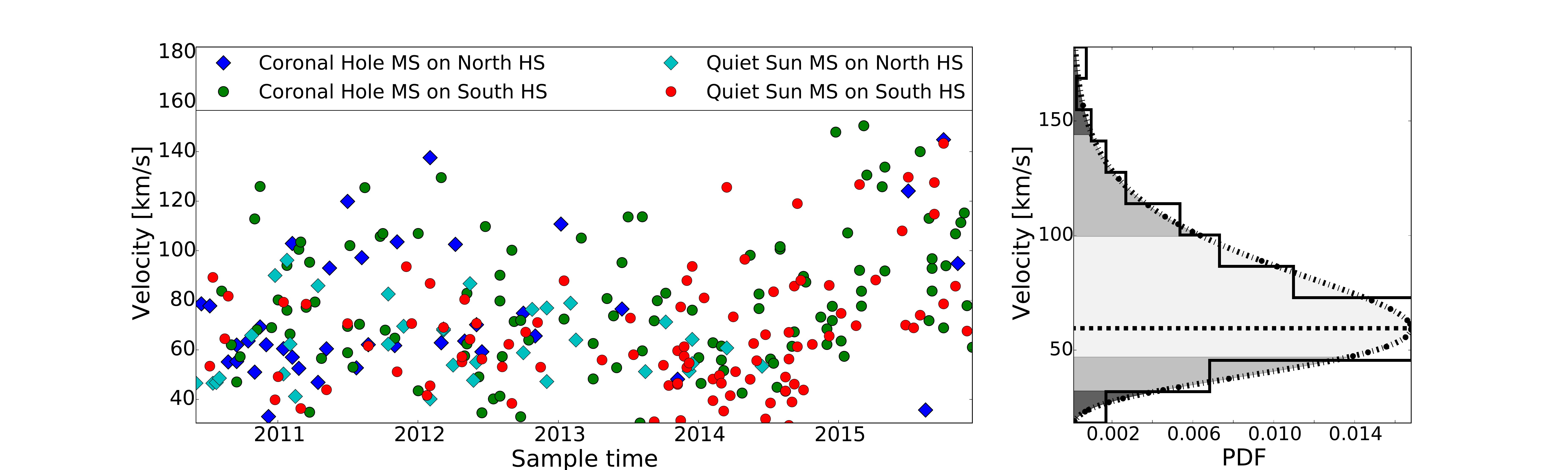}
	\includegraphics[scale=0.2]{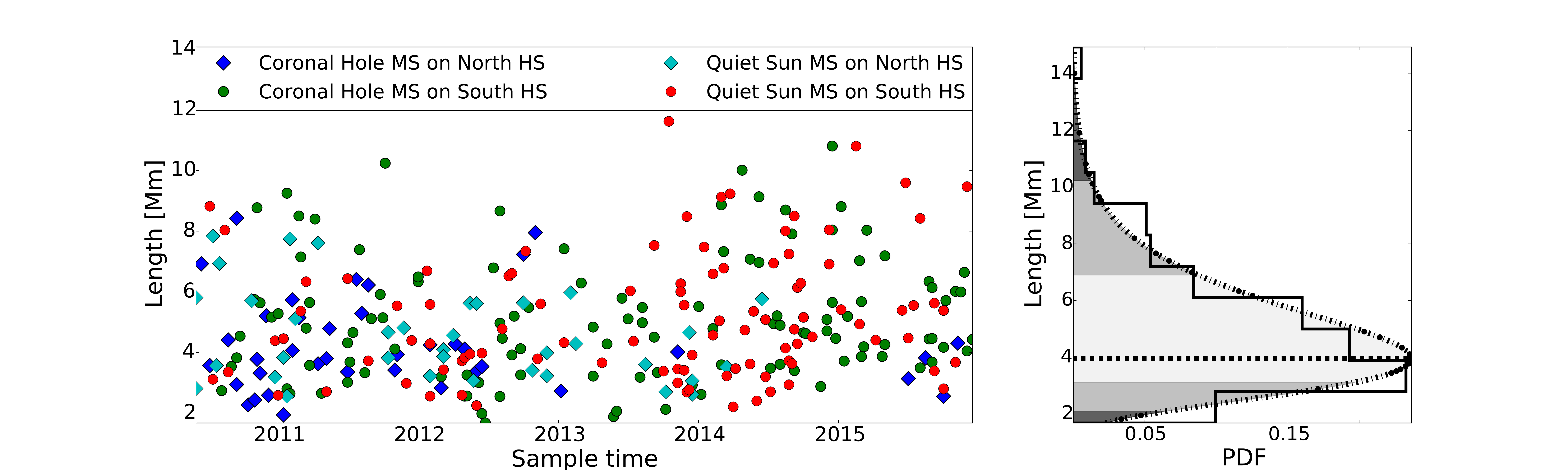}
	\caption{Distributions, from top to bottom: maximum area, maximum length, lifetime, average velocity and maximum width. In each panel, on the LHS, the distribution of the actual property can see. Different types of mark are used for different hemispheres: diamonds represent the northern hemisphere (dark blue -- CH, light blue -- QS), circles denote the south (green -- CH, red -- QS). On the RHS panels the histogram of each distributions is provided. Dashed line indicates the fitted log-normal distributions. The vertical line represents the mode of the distribution (light-darker-dark strips correspond to 1$\sigma$, 2$\sigma$, 3$\sigma$ distributions).}
	\label{histdist}
\end{figure*}

\cite{kayshap2013} carried out a detailed description of the evolution of an individual MS. This jet occurred in the north polar corona on 2010 November 11 with nearly 24 minutes lifetime and 40 Mm height. Based on its detailed observational description, a two-dimensional numerical simulation was performed using the VAL IIIC initial atmospheric  model \citep{vernazza1981}. A small-scale magnetic flux tube emerged from the sub-photospheric layers into the previously existed open magnetic field, than underwent on kink-oscillation. The authors claimed that this kinking motion is the source of the MS. Flux-emerging model is one of the widely accepted theory of solar jets (for review of this topic, see \cite{sterling2000}).

According to the study by \cite{gyenge2015}, using SDO/AIA data, the spatial distribution of macrospicules shows inhomogeneous but systematic properties. It is known that the latitudinal distribution of MS concentrates around the poles. However, from the study of \cite{gyenge2015} the longitudinal distribution also shows inhomogeneous properties. The longitudinal positions tend to focus around certain belts which suggests that there is a relationship between the position occurrence and the generation of the global magnetic field.

Taking advantage of the uninterrupted observations of SDO, \cite{bennett2015} built up a dataset with 101 MS covering observation of a 2.5 year-long time interval. The authors claim that features like maximum length ($28$ Mm), upflowing velocity ($110 \text{ km s}^{-1}$) and lifetime ($14$ min) are varying in time systematically. Cross-correlation of maximum length-maximum velocity and lifetime-maximum length showed a significant correlation ($k=0.43, 0.66$), however the correlation coefficient is much smaller in the case of maximum velocity and lifetime ($k=0.16$). Ballistics of MS were investigated as well, which indicates a strong influence of gravity in the rise and fall of MS. Taking stock of the ballistics and two characteristic density values of MS ($\rho = 1.0 \times 10^{−8} \text{kg m}^{−3}$ and $\rho= 1.0 \times 10^{−11} \text{kg m}^{−3}$), the authors estimated the formation energy $1.46 \times 10^{17} \text{J}, 4.78 \times 10^{16} \text{J}, 3.09 \times 10^{15} \text{J}, 1.46 \times 10^{14} \text{J}$ for 25, 5, 1, and 0.5 Mm scale heights and $3.66 \times 10^{13} \text{J}$ for uniform plasma distribution cases.

In this study, we focus on the statistical investigation of the temporal behaviour of MS on timescales spanning considerably longer than previous studies, i.e. for a time interval of just under 6 years. This extended temporal investigation yields the opportunity to analyze the temporal variation of MS properties and the detection of the oscillatoric behavior of their properties on about by-annual time scales, well-beyond the life-time of individual MS. In Section~\ref{database}, we introduce the way the database was built up. In Section~\ref{results}, we discuss the results of the statistical analysis. We outline our discussion and conclusions in the last section.

\section{Database}
\label{database}
The source and driving force of majority of solar phenomena is the large-scale, global magnetic field, which varies on a long-term, e.g. 11-year time period. However, the temporal behaviour of jets is always considered in short time-scale. That may be the reason, why previous studies investigated only a few MS over a short time period. The temporal ranges of jet evolution are much shorter than the characteristic timescale of the solar cycle. Therefore, investigation of the properties of short time-scaled macrospicules over a long-time time period is still an uncharted territory with some interesting questions to answer to. To achieve this objective, an instrument is required i) to have high temporal and spatial resolution to properly resolving the jets themselves, and ii) to operate continuously for multiple years in order to investigate long-term evolution in statistical sense. The ideal choice is SDO, which was launched in June 2010, near to solar minimum between Solar Cycles 23 and 24. The operation of SDO for a period of about 6 years, as of writing, provides a great opportunity to acquire the long-term temporal variation of the properties of MS in regards to the current Solar Cycle.
\begin{figure*}[!t]
	\centering
	\includegraphics[scale=0.42]{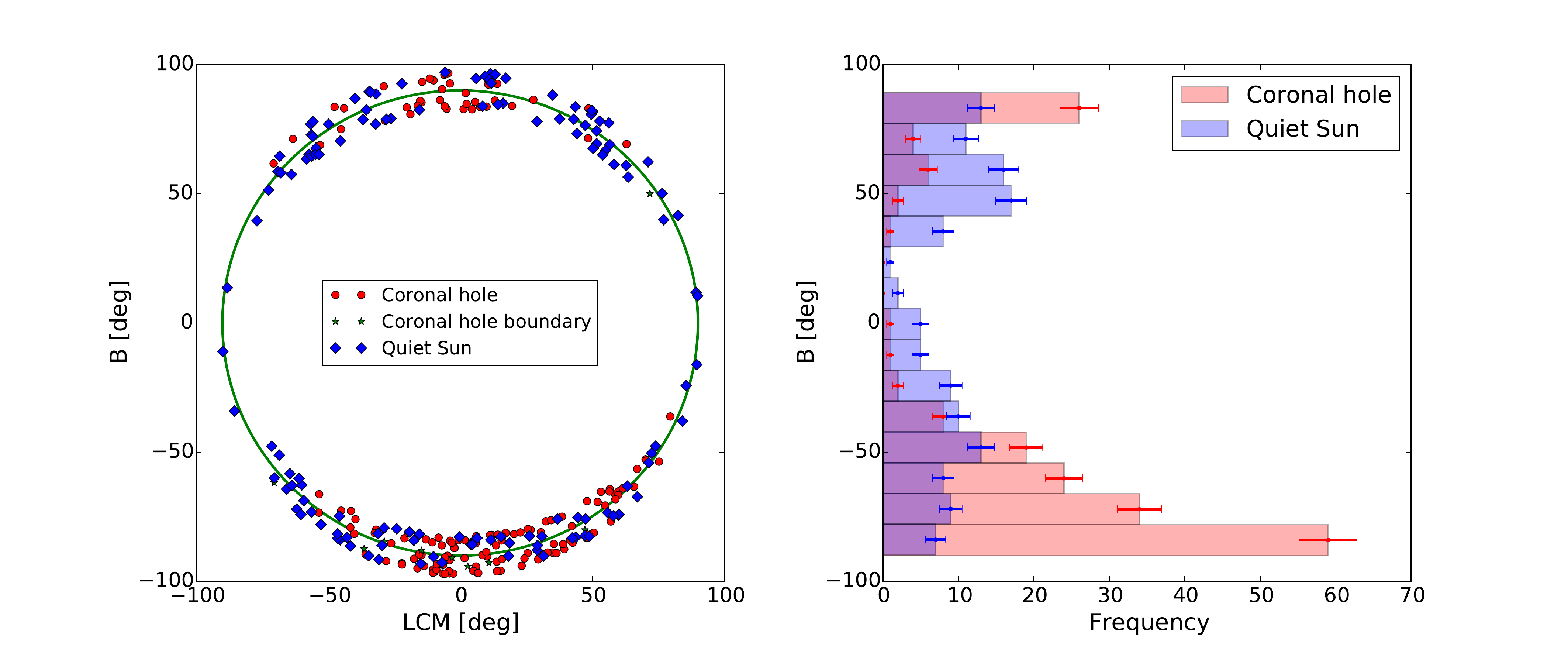}
	\caption{On the left-hand side locations of all investigated MS are plotted around the solar limb. Red, blue and green marks the Coronal Hole, Quiet Sun and Coronal Boundary MS, respectively. The vertical axis is the heliographic latitude (B). Longitude of Central Meridian (LCM) is along the horizontal axis. On the right-hand side, histogram of CH-MS and QS-MS location is provided with red and blue strips, respectively.}
	\label{spatdist}
\end{figure*}

\begin{table*}[htbp]
	\caption{Preferences of the fitted log-normal distributions}
	\centering
	\begin{tabular}{|l|cccccc|}
	\hline
	\multirow{2}{*}{ } & \multirow{2}{*}{\textbf{Mode}} & \multirow{2}{*}{\textbf{Mean}} & \multirow{2}{*}{\textbf{Median}} & \multicolumn{3}{c|}{\textbf{Distribution}} \\
	& & & &1 $\sigma$ & 2 $\sigma$ & 3 $\sigma$\\
	\hline  
	Maximum length [Mm] & 24.95 & 28.07 & 26.99 & 20.39 -- 35.72 & 15.41 -- 47.27 & 11.64 -- 62.56\\
	Maximum lifetime [s] & 916.72 & 1015.88 & 981.69 & 755.68 -- 1275.3 & 581.70 -- 1656.73 & 447.77 -- 2152.24\\ 
	Maximum width ["] & 3.95 & 4.98 & 4.61 & 3.11 -- 6.83 & 2.10 -- 10.12 & 1.42 -- 14.99\\
	Average velocity [km/s] & 59.62 & 73.25 & 68.39 & 47.22 -- 99.05 & 32.60 -- 143.46 & 22.51 -- 207.77\\
	Maximum area [Mm$^2$] & 69.01 & 97.787 & 87.06 & 53.77 -- 140.97 & 33.2 -- 228.26 & 20.5 -- 369.61\\
	\hline
    \multicolumn{7}{c}{\textbf{Note:} 1$\sigma$ distribution covers 68\%, 2$\sigma$ represents 95\%, 3$\sigma$ marks 99.5\% of data} 	
	\end{tabular}
	\label{hist}
\end{table*}
SDO/AIA 30.4 nm images may be divided into three regions due their overall average intensity. Active Regions [AR] are bright, fine-detailed areas often near the solar equator; at the same time coronal holes [CH] are mostly formed around the solar poles and their reduced intensity is related to the open magnetic field \citep{aschwanden2004}. Finally, Quiet Sun area [QS] is defined as non-AR or non-CH solar surface. Therefore, the macrospicules we studied were named and catalogued as Coronal Hole [CH-MS] and Quiet Sun [QS-MS] macrospicules, distinguished by the surrounding solar environment. We have not investigated AR MS, for reasoning see below.

To build a long-term database with sufficient amount of observations for statistical analysis, a strict definition of what is actually considered a MS is necessary. Let us summarize how we define an MS, by the following five points. First, MS are "thin" jet phenomena at the solar limb. Second, MS generated in Active Regions are not selected, just those from Coronal Hole and Quiet Sun areas. The reason for this apparently perhaps too restrictive criteria is that the large-scale magnetic field of ARs is able to drive MS-like phenomena \citep{chandra2015,sterling2016}, which could be different in terms of physics from MS formed in CH or QS areas \citep{kayshap2013}. Third, MS are shorter than 200 pixels (70 Mm). This, perhaps somewhat arbitrary, upper limit avoids to contaminate our dataset by some other high-energetic jet phenomena. Fourth, MS have to have a visible connection with the solar surface. The lack of connection may mean that the MS is formed on the "other side", i.e. far side, of the solar limb. If far-side MS were considered, this may carry an error to estimate the distance between the top and the footpoint of MS. Fifth, and perhaps most distinctly, 1-2 minutes before the appearance of the MS jet, a brightening is identifiable at the solar surface. The presence of brightening actually also confirms that the MS is on the visible side of the Sun. A number of previous works discussed the physics of brightening generation \citep{pike1997, sterling2015}. 

To provide a temporally homogeneous dataset, MS were identified and chosen from a period of 2-hour long interval between 12:00 and 14:00 on every 1$^{\text{st}}$, 7$^{\text{th}}$, 15$^{\text{th}}$, 24$^{\text{th}}$ days of each months from 01/06/2010 until 12/31/2015.

To turn macrospicules identified by SDO observations to geometrically, e.g. morphologically, measurable features, MS were fitted with tetragons. Note, we are not saying MS have a tetragon shape. On contrary, MS have an irregular shape. We only approximate their geometric extent in order to estimate their length, width, etc. A framework of Sunpy (see e.g. \citealt{mumfort2013}) was used for this process. On every tetragon, two of the diagonal points represent the bottom and top and the distance between them measures the actual length of the jet. The distance between the remaining two points model the width of MS perpendicular to the main axis. Furthermore, the position of bottom point of MS was assigned with the actual observing time. Data about a MS contains the position of four vertices for every frame of observation, where the MS is clearly visible.

By applying the above outlined criteria to SDO/AIA 30.4 nm observation, 301 MS have been detected during 5.5 years of sequence of observation. The number of different types of macrospicule are not equal in terms of location, 158 jets formed ($\approx$ 52.48\% of all MS) in Coronal Holes and only 134 took place ($\approx$ 44.50\% of all MS) in Quiet Sun regions. Occasionally a MS was not registered into this dual system, therefore a new class, named Coronal Hole Boundry [CHB], was assigned to catalogue them. However the number of CHB macrospicules is only 9 ($\approx$ 0.02\% of all MS). That is the reason, why we have not applied to CHB MS the same statistical studies like we have in the case of CH or QS macrospicules. 

On-disk MS are often associated with explosive phenomena. These events (therefore MS as well) are regions of excess line width, which were observed recently in greater extent in Coronal Holes than Quiet Sun regions \citep{kayshap2015}.
\section{Results}
\label{results}
\subsection{Spatial distribution on solar disc}
For each frame during the entire lifetime of a MS, the position of the four edges of their tetragon is built up from two polar coordinates: the first represents the distance from the solar disc center, the second one shows a polar degree from along the solar limb. From these two data values the actual position of MS on the solar disc is easily determinable. Because each tetragon was fitted individually at different time frames, the position of the associated MS varies on a small, approximately 3 arcsec distance. This variation in neglectable, so in this study, the position of MS is always denoted with the position of the brightening observed on the first frame. 
\begin{figure*}[t]
	\centering
	\includegraphics[scale=0.3]{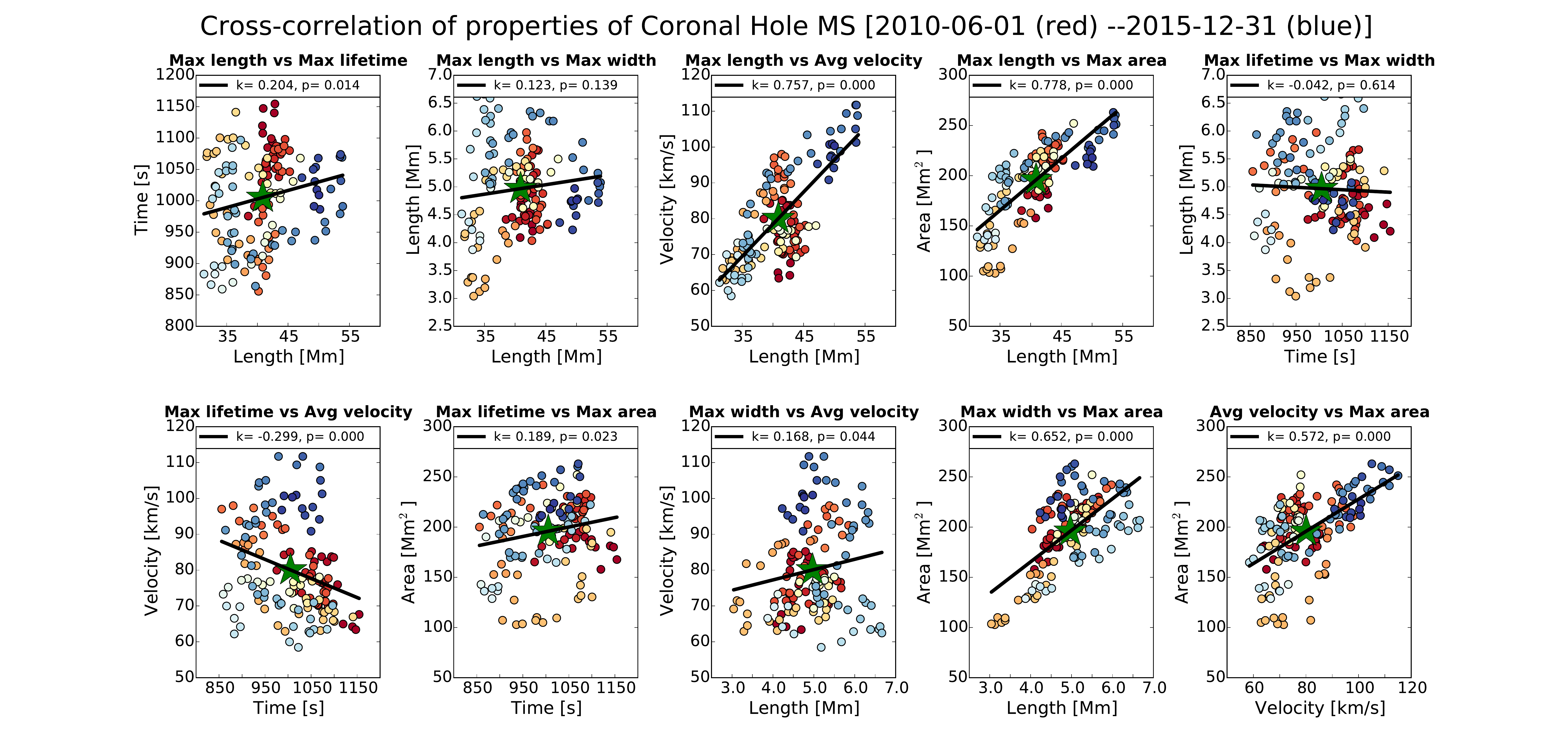}
	\includegraphics[scale=0.3]{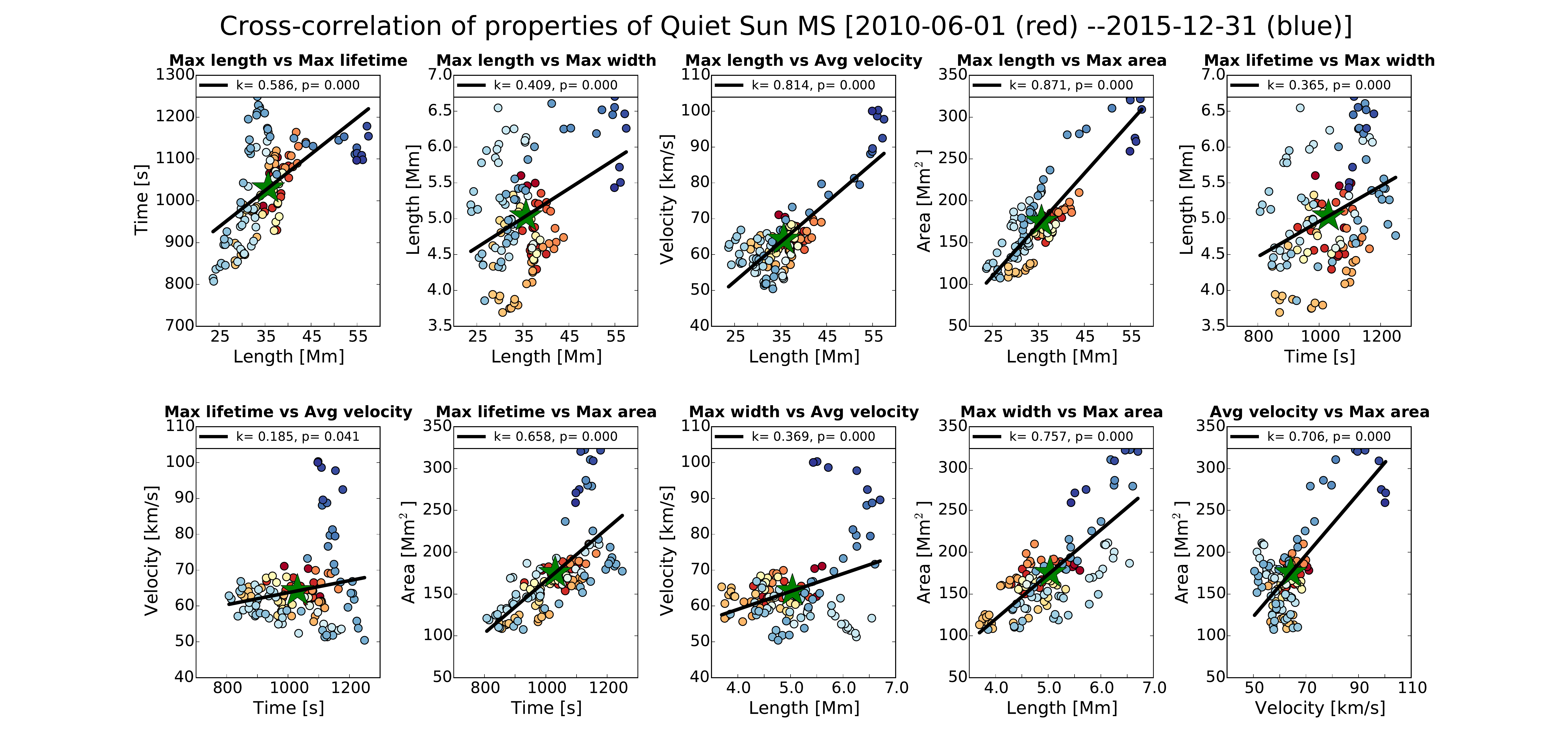}
	\caption{Cross-correlation of a range of MS properties. Variation in color of markers represents the progress in time: red indicates June 2010, blue indicates December 2015. Green star indicates the weighted geometric center of property for a given plot. Characteristics of a fitted black linear are the $k$ correlation coefficient and two-sided \textit{p}-value obtained in a hypothesis test.}
	\label{crosscorr}
\end{figure*}

By estimating the positions of all the observed MS in our database, it is clearly visible that MS are formed mostly around the solar poles. There may be a bias caused by the introduced selection criteria, namely we exclude ARs that are more abundant around the solar equator. Further, the ratio between the number of CH-MS and QS-MS could be influenced by the solar dynamo: when the poloidal field is more dominant (e.g. during solar minima), the area of coronal hole regions may be larger. Therefore the number of CH-MS could grow substantially. This effect may be reversed during solar maxima with less dominant poloidal fields.
The histogram of the spatial distribution shows, what is expected: CH-MS mostly take place around the solar poles, while QS-MS cuddle around them as a "ring" as seen in the right-hand side of Figure~\ref{spatdist}. 

Another interesting aspect is the ratio between the number of MS on the two solar hemispheres. For QS-MS, the corresponding numbers are nearly equal ($n_{\text{North\_QS}}=69$, $n_{\text{South\_QS}}=65$), but a huge difference is found between the northern and southern CH-MS numbers ($n_{\text{North\_CH}}=39$, $n_{\text{South\_CH}}=119$). 

Strong asymmetry between the two hemispheres is visible in the heliospheric current sheet position, therefore the inclination of polar jets \citep{nistico2015}. For the period of 2007 -- 2008 solar minima, the authors found that jets are deflected towards low latitudes and this deflection is more pronounced
at the north pole than at the south pole. Asymmetry was also reported in many different magnetic solar phenomena: e.g. sunspots \citep{chowdhury2013, kitchatinov2014}, global distribution of the solar wind speed \citep{hoeksema1995} and magnetic field measurements in the interplanetary medium \citep{erdos2010}. These results suggest an influencing effect by the solar dynamo \citep{shetye2015}.
\begin{figure*}[!t]
	\centering
	\includegraphics[scale=0.3]{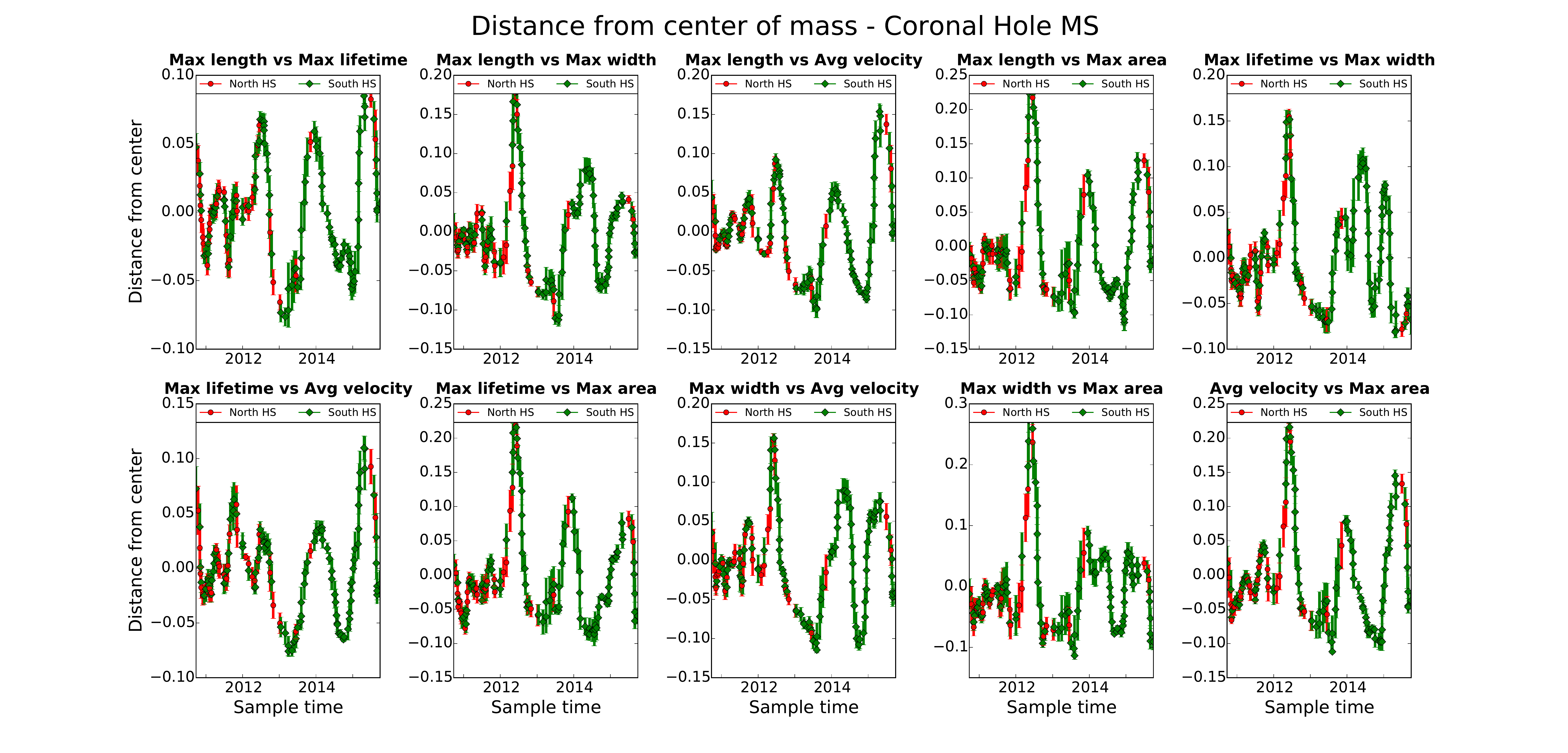}
	\includegraphics[scale=0.3]{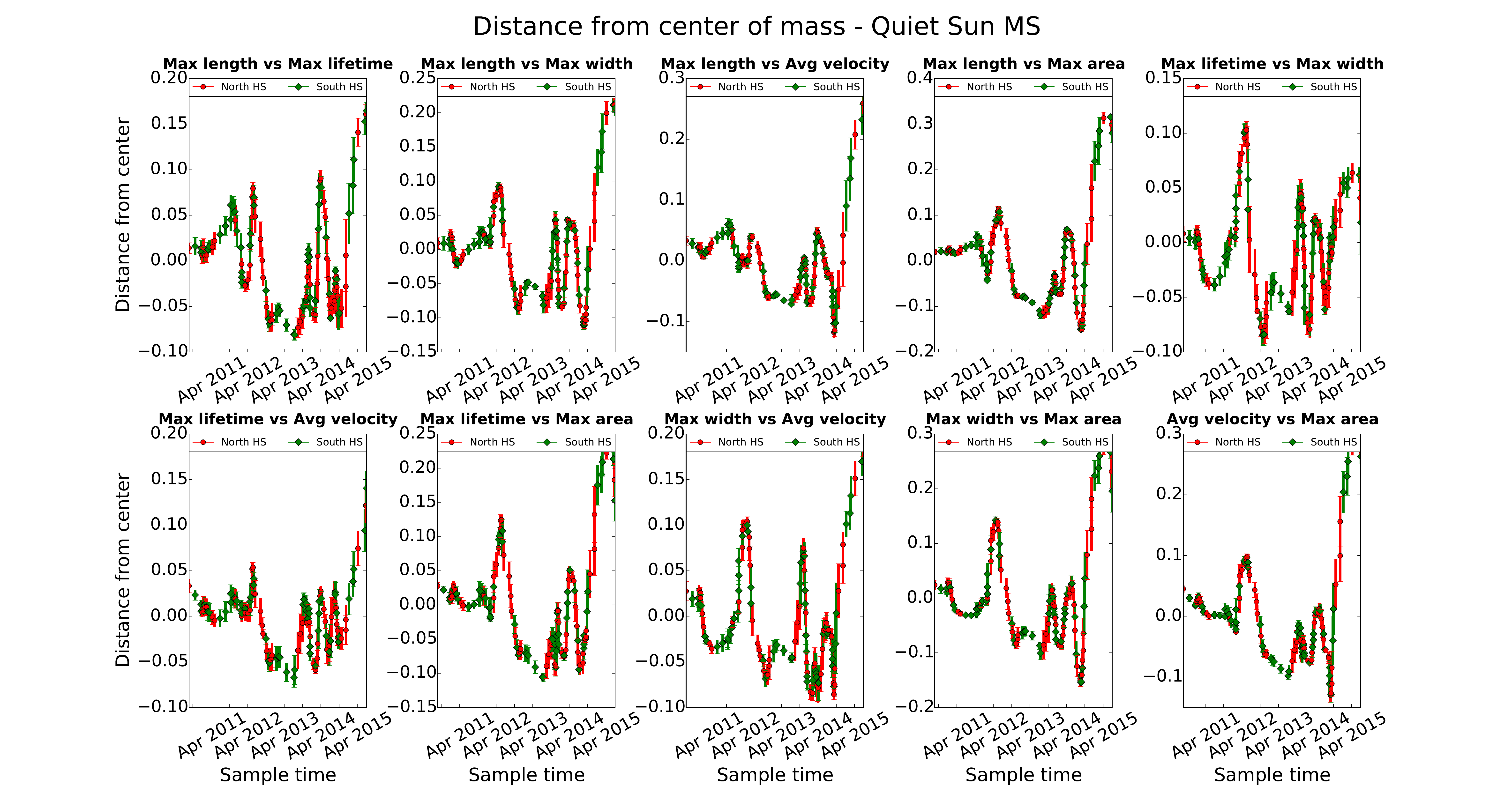}
	\caption{Distance plot of each marks in Figure~\ref{crosscorr} from the center of mass. Red points indicates the north hemisphere values, green marks represent MS from south hemisphere}
	\label{distplot}
\end{figure*}

\subsection{Measuring macrospicule properties}
Many different properties of MS, such as length, lifetime, width, area and rise (i.e. emergence) velocity, have been refined in several studies in the last 40 years. Therefore a crucial cross-checking point of this study now is how well our results fit with those reported by others. The tetragon assumption is a powerful tool to determine key properties in the following way: a) maximum length is the greatest distance between the bottom and the top points; b) maximum width represents the greatest distance between the two side diagonal points; c) average upflow velocity is the average of speed values calculated from spatial difference between two frames in the emerging epoch of MS life; d) lifetime is the temporal difference between the first and last frame, where the MS is visible e) maximum area shows the maximal value of geometrical areas of all tetragon during the whole lifetime of a MS.
    
As a first approximation, determining these values for each MS and plotting them in time, no obvious temporal variation is found. To have a more accurate understanding of the range of characteristic values determining of MS properties, their histogram was investigated. A log-normal distribution is fitted to each histogram as seen in Figure~\ref{histdist}. The main properties of each fit (e.g. mode, median, mean, scattering, distribution) to the data are in Table~\ref{hist}. These obtained values are found to be in line to those reported previously in the literature. In particular, three properties show great similarity: lifetime ($16.75 \pm 4.5$ minutes), maximum width ($6.1 \pm 4$ Mm) and average upflow velocity ($73.14 \pm 25.92$ km s$^{-1}$). The maximum length is nearly overestimated here, with $28.05 \pm 7.67$ Mm. These similarities indicate a positive feedback to the criteria, that we have set for defining MS, and, the applicability of tetragon assumption, as well. For a summary of comparison of the average properties found here with those reported in earlier studies, see Table~\ref{lit}.

\begin{table*}[htbp]
	{\footnotesize  
	\caption{Summery of macrospicules based on previous studies}
\begin{center}
	\begin{tabular}{|l|rrrrr|}
	\hline
	${}$ & \cite{bohlin1975} & \cite{labonte1979} & \cite{dere1989} & \cite{bennett2015} & New results (1 $\sigma$)\\
	\hline
	Length [Mm] & 6.24--19.45 & 6.24-25.67 & 3.9-17.9 & 14-68.46 & 20.39-35.72\\
	Lifetime [min] & 8-45 & 4-24 & $\geq$ 3 & 2.7-30.6 & 12.5-21\\
	Velocity [km/s] & 10-150 & $\leq$ 60 & 20-50 & 54.1-335.5 & 47.22-99.05\\
	Width [Mm] & 3.6-10.9 & 2.2-6.5 & - & 3.1-16.1 & 2.1-10.1\\
	Spectral line & He II ($\lambda 304$) & H$\alpha$ ($\lambda 6562$) & C IV ($\lambda 1548$) & He II ($\lambda 304$) & He II ($\lambda 304$)\\
	$\log$ temperature of line [K] & 4.9 & 4.0 & 5.0 & 4.0 & 4.0\\
	Number of MS & $\sim$ 25 & 32 & 10 & 101 & 301\\
	Spatial resolution ["] & 2 & 0.5 & 2 & 1.5 & 1.5 \\
	Temporal resolution [s] & $\geq$ 180 & $\sim$ 60 & 20, 60 & 12 & 12\\
	\hline
	\end{tabular}
	\label{lit}
	\end{center}}
\end{table*}

\subsection{Cross-correlation of properties}
In a next step, we now propose to aquire, whether there is any  correlation between the properties of MS found. For this reason, selected properties are plotted against each other. To show whether any two properties correlate, a linear fitting is applied to the data distribution. This fit is displayed in Figure~\ref{crosscorr}. The correlation coefficients ($k$) were determined also, which seems to have a relatively small value for both CH ($k_{\text{CH}}=[-0.29, -0.012, 0.127, 0.131, 0.195, 0.214, 0.557]$) and QS ($k_{\text{QS}}=[0.179, 0.355, 0.373, 0.406, 0.58]$) type MS. This suggests, that there is no strong correlation between the different properties of MS in general. In both cases, only a small number of combinations show stronger correlations. "Maximum length vs Maximum area" ($k_{\text{CH}}= 0.78, k_{\text{QS}}=0.87$), "Maximum width vs Maximum area" ($k_{\text{CH}}= 0.64, k_{\text{QS}}=0.75$) and "Maximum length vs Average velocity" ($k_{\text{CH}}= 0.76$, $k_{\text{QS}}=0.81$) seem to have relatively strong correlation for both CH-MS and QS-MS. A correlation between datasets cannot be negligible if the correlation coefficient is greater than around $0.6$. In this respect the above stronger correlations seem to be logical, as area data are derived from the length and width. Further, "Maximum area vs Average velocity" ($k_{\text{QS}}=0.71$) and "Maximum area vs Maximum lifetime" ($k_{\text{QS}}=0.65$) pairs indicate a stronger correlation for QS-MS only. 

Next, there is a visible gap between CH-MS and QS-MS in their average correlation coefficient ($|k_{\text{CH\_avg}}|=0.372, |k_{\text{QS\_avg}}|= 0.57$) and the ratio of the correlation coefficient is also greater than 0.6 ($n_{\text{k\_CH}}=3/10$, $n_{\text{k\_CH}}=5/10$). The source of these differences may root in the underlying governing physical differences between the two types of MS, which should be clarified in a future research, that is not within the scope of the present paper.
 
By enquiring the large-scale time-dependence, i.e. solar cycle evolution, in this study, an interesting effect is visible. Namely, in Figure~\ref{crosscorr}, the change of marker color represents the march of time. Red marks shows the value of an MS property around June 2010 and the sequential color variation into blue captures the progress in time. By following the conversion of color from red to blue leads to a trajectory in each figure. In some aspects, at least conceptually, these paths are similar to different branches in the well-known Hertzsprung-Russell diagram (HRD). To characterize these paths in time, the entire problem is simplified to a geometrical problem. Fix points of these distributions became the geometrical center of "mass", which are marked by a green star in each panel of Figure~\ref{crosscorr}. 
To give a correct distance of each points from this center point, all the properties are normalized due the different magnitudes of values (e.g. lifetime values are three magnitudes greater, than maximum width values). After normalization, the geometric distance can be easily worked out. As the cross-correlation distributions suggested, strong temporal variation is visible in all cases in Figure~\ref{distplot}. Multiple strong oscillations are visible in these plots on short timescale ($\tau_{\text{osc}} <$ 2 years). Further, the characteristics of the obtained distance curves are different between CH and QS macrospicules, but there is a dominant peak around the first half of 2012 in the majority of the distance plots. All of these suggest that there is a deterministic underlying temporal variation in the raw data, which was not known earlier to the best of knowledge of these authors. Future research of the oscillatory pattern may found here connect macrospicules to oscillatory behaviour found in other, large-scale solar phenomena, already reported to show a quasi-biennial oscillation \citep{fletcher2010, broomhall2015} or oscillations with even shorter periods, e.g. found by \cite{cho2013, gyenge2014, gyenge2016}.

\section{Conclusion}
\label{conclusion}
The aim of this study is to yield a more accurate and precise statistically relevant characteristic study of the physical properties of macrospicules. For this reason, a dataset is built by means of analysing 301 MS over a 5.5 year-long time interval between June 2010 and December 2015. The raw data were obtained by SDO/AIA using the 30.4 nm wavelength band, where the MS jets were detected at the solar limb. The underlying fundamental principle of this database is to form and apply a range of criteria to MS, which can be summarized into five points. In order to gather information about each MS, which fits well the required criteria at every frame of their lifetime, these jets are fitted with tetragons. The diagonals between the edges of the tetragon represents the physical dimensions of the MS such as length and width. 

Five observed properties of MS were analysed in this study: maximum length, maximum width, average upflow velocity, maximum area, lifetime.

Taking advantage of this dataset, the spatial distribution of some key parameters of MS was investigated first. Because Active Regions were excluded from the solar area of investigation by definition, where MS could be identified, the observed jets are found to be at high(er) solar latitudes. CH jets accumulate around the solar poles due to the large-scale open magnetic field of polar Coronal Holes. Quiet Sun MS form a "ring" around them. Further, a strong asymmetry is visible in the number of CH-MS between the hemispheres. The number of CH-MS on the southern hemisphere is almost three times larger, than that of the northern hemisphere. The source of this difference may be the solar dynamo as that was shown for e.g. sunspot area \citep{chowdhury2013, kitchatinov2014}. Future research in this topic should seek for a connection between asymmetry indexes of multiple magnetic structures.

Distributions of MS properties in time show a strong temporal variation. To have accurate estimates for the average properties, the temporal variation was put aside and their histograms were studied initially. Each histogram was fitted with a log-normal distribution and their preferences (e.g. mode, mean, median, distribution) were determined to characterise the MS. Comparing our findings to those of the previous studies (note, that values here are 1$\sigma$ distribution of the log-normal distributions) we conclude that: lifetime ($16.75 \pm 4.5$ min), width ($ 6.1 \pm 4 $ Mm), average velocity ($73.14 \pm 25.92$ km s$^{-1}$) values are in an agreement, while length ($28.05 \pm 7.67$ Mm) is slightly greater. Correspondence between the results found in this research and those in the literature verify the choice of the tetragon assumption. 

Last but not least, cross-correlation of the raw data was investigated. Fitting a linear and cross-correlated, there is often a lack of a dominant correlation ($|k_{\text{CH\_avg}}|=0.372, |k_{\text{QS\_avg}}|= 0.57$). In three cases, however, the coefficients are found to be relatively strong for both types of MS. For QS-MS, further two combinations of physical parameters (e.g. "Maximum area vs Average velocity" and "Maximum area vs Maximum lifetime") show a stronger correlation, which may reveal the underlying physical differences between the formation of CH-MS and QS-MS. Considering the temporal evolution of these distributions, remarkable paths became visible in cross-correlation visualisations. To study this behaviour, the distances between the center of mass and each points were determined. These distance plots in time reveal a strong, previously unseen temporal variation in the database. 

\cite{broomhall2009} found that variation of the frequency shift of the global $p$-mode oscillation is a superposition of two oscillators: a stronger one, which has the well-known 11-years long oscillation and a weaker one, where the period is around two years. These oscillations are named as quasi-biennial oscillations (QBOs) and discovered first by \cite{belmont1966}. QBQs were found in many {\it global} solar phenomena. \citet{penza2006} found that the reconstructed dataset of the line depth of three photospheric line over 25 years shows a QBO. \citet{zaqarashvili2010} investigated the stability of magnetic Rossby waves in the solar tachocline and their results indicate a Rossby wave harmonic with a period of $\sim$2 years, a possible source of QBOs. 

\cite{fletcher2010} suggests that the source of these oscillations could be a second dynamo layer near to the solar surface. \citet{broomhall2015} also found QBOs in measuring proxies of the magnetic field in the Sun. Lately, \cite{beaudoin2016} constructed a double dynamo-layer model, which is able to excite QBOs. If {\it localised} solar features like MS, by means of statistical investigation of their properties during solar cycle time scales show a similar behaviour, that would suggest a connection between {\it local} dynamics (e.g. MS) and the evolution of \textit{global} magnetic field (e.g. solar cycle), an unrevealed question with great potentials. Therefore, this will be the focus of our follow-up research.

\section{Acknowledgments}

\small{The authors thank S. Bennett for helpful advice at the beginning of this research. The authors acknowledge the support received from ESPRC (UK) and the Erasmus Programme of EU for enabling this research. All the results are derived using Python, an open-source and community-developed programming language and its solar data analysis package, Sunpy \citep{mumfort2013}. RE is grateful to STFC (UK), The Royal Society and the Chinese Academy of Sciences President’s International Fellowship Initiative, Grant No. 2016VMA045 for support received.}


\newpage
\begin{thebibliography}{99}

\bibitem[\protect\citeauthoryear{Aschwanden}{2004}]{aschwanden2004} Aschwanden, M.; 2004; Physics of the Solar Corona; Springer-Verlag Berlin Heidelberg

\bibitem[\protect\citeauthoryear{Beckers}{1968}]{beckers1968} Beckers, J. M.; 1968; Solar Physics; 3:367--433

\bibitem[\protect\citeauthoryear{Beaudoin et al.}{2016}]{beaudoin2016} Beaudoin, P.;  Simard, C.; Cossette, J.-F.; Charbonneau, P.; 2016; The Astrophysical Journal; 826:138--151

\bibitem[\protect\citeauthoryear{Belmont et al.}{1966}]{belmont1966} Belmont, A. D.; Dartt, D. G.; Ulstad, M. S.; 1966; Journ. of Atm. Sci.; 3:314--319 

\bibitem[\protect\citeauthoryear{Bennett \& Erd\'{e}lyi}{2015}]{bennett2015} Bennett, S. M.; Erd\'{e}lyi, R.; 2015, The Astrophysical Journal, 808:135--144

\bibitem[\protect\citeauthoryear{Broomhall et al.}{2009}]{broomhall2009} Broomhall, A.-M.; Chaplin, W. J., Elsworth, Y., et al; 2009; The Astrophysical Journal; 700:162--165

\bibitem[\protect\citeauthoryear{Broomhall \& Nakariakov}{2015}]{broomhall2015} Broomhall, A.-M.; Nakariakov, V. M.; 2015, Solar Physics, 290:3095--3111

\bibitem[\protect\citeauthoryear{Bohlin et al.}{1975}]{bohlin1975}  Bohlin, J. D.; Vogel, S. N.; Purcell, J. D.; et al.; 1975; The Astrophysical Journal; 197:133-135

\bibitem[\protect\citeauthoryear{Chandra et al.}{2015}]{chandra2015} Chandra, R.; Gupta, G. R., Mulay, S.; Tripathi, D.; 2015; MNRAS, 446:3741--3748

\bibitem[\protect\citeauthoryear{Chowdhury et al.}{2013}]{chowdhury2013} Chowdhury, P.; Choudhary, D. P.; Gosain, S.; 2013; The Astrophysical Journal; 768:188--198

\bibitem[\protect\citeauthoryear{Cho et al.}{2013}]{cho2013} Cho, I.-H.; Hwang, J.; Park, Y.-D.; 2013; Solar Physics; 289:707--719

\bibitem[\protect\citeauthoryear{de Pontieu et al.}{2012}]{depontieu2012} de Pontieu, B.; Carlsson, M.; Rouppe, L. H. M.; et al.; 2012; The Astrophysical Journal; 752:12--18

\bibitem[\protect\citeauthoryear{de Pontieu et al.}{2004}]{depontieu2004} de Pontieu, B.; Erd\'elyi, R.; James, S. P.; 2004; Nature; 430:536--539

\bibitem[\protect\citeauthoryear{de Pontieu et al.}{2007}]{depontieu2007} de Pontieu, B.; McIntosh, S.; Hansteen, V. H.; et al.; 2007; Publications of the Astronomical Society of Japan; 59:655--662

\bibitem[\protect\citeauthoryear{Dere et al.}{1989}]{dere1989} Dere, K. P.; Bartoe, J.-D. F.; Brueckner, G. E.; et al.; 1989; Solar Physics; 119:55--63

\bibitem[\protect\citeauthoryear{Erd\H{o}s \& Balogh}{2010}]{erdos2010} Erd\H{o}s, G.; Balogh, A.; Journal of Geophysical Research: Space Physics; Volume 115; Issue A1; CiteID A01105

\bibitem[\protect\citeauthoryear{Fleck et al.}{2000}]{fleck2000} Fleck, B.; Brekke, P.; Haugan, S.; Duarte, L.; et al.; 2000; ESA Bulletin; 102:68--86

\bibitem[\protect\citeauthoryear{Fletcher et al.}{2010}]{fletcher2010} Fletcher, S. T.; Broomhall, A.-M.; Salabert, D.; et al.; 2010; The Astrophysical Journal Letters; 718:19--22

\bibitem[\protect\citeauthoryear{Gyenge et al.}{2014}]{gyenge2014} Gyenge, N.;  Baranyi, T.; Ludm\'any, A.; 2014; Solar Physics; 289:579--591


\bibitem[\protect\citeauthoryear{Gyenge et al.}{2015}]{gyenge2015} Gyenge, N.; Bennett, S.; Erd\'{e}lyi, R.; 2015; Journal of Astrophysics and Astronomy; 36:103--109

\bibitem[\protect\citeauthoryear{Gyenge et al.}{2016}]{gyenge2016} Gyenge, N.; Ludm\'any, A.; Baranyi, T.; 2016; The Astrophysical Journal; 818:127--135

\bibitem[\protect\citeauthoryear{Hoeksema}{1995}]{hoeksema1995} Hoeksema, J. T.; Space Science Reviews; 72:137-148

\bibitem[\protect\citeauthoryear{Judge}{2006}]{judge2006} 
Judge, P.; 2006; Solar MHD Theory and Observations: A High Spatial Resolution Perspective ASP Conference Series; 354:259

\bibitem[\protect\citeauthoryear{Kayshap et al.}{2015}]{kayshap2015} 	
Kayshap, P.; Banerjee, D.; Srivastava, A. K.; 2015; Solar Physics; 290:2889--2908

\bibitem[\protect\citeauthoryear{Kayshap et al.}{2013}]{kayshap2013} 	
Kayshap, P.; Srivastava, A. K.; Murawski, K.; Tripathi, D.; 2013; The Astrophysical Journal; 770:3--11

\bibitem[\protect\citeauthoryear{Kitchatinov \& Khlystova}{2014}]{kitchatinov2014} 	
Kitchatinov, L. L.; Khlystova, A. I.; 2014; Astronomy Letters: A Journal of Astronomy and Space Astrophysics; 40:663--666

\bibitem[\protect\citeauthoryear{Kuridze et al.}{2015}]{kuridze2015} 	
Kuridze, D.; Henriques, V.; Mathioudakis, M.; et al.; 2015; The Astrophysical Journal; 26--34

\bibitem[\protect\citeauthoryear{Labonte}{1979}]{labonte1979} Labonte, B. J.; 1979; Solar Physics; 63:283--296

\bibitem[\protect\citeauthoryear{Lee \& Deane}{2009}]{lee2009} Lee, D.; Deane, A. E.; 2009; J. Comput. Phys.; 228:952

\bibitem[\protect\citeauthoryear{Lemen et al.}{2012}]{lemen2012} Lemen, J. R.; Title, A. M.; Akin, D. J.; et al.; 2012; Solar Physics; 275:17--40

\bibitem[\protect\citeauthoryear{Lipartito et al.}{2014}]{lipartito2014} Lipartito, I.; Judge, P. G.; Reardon, K.; Cauzzi, G.; 2014; The Astrophysical Journal; 785:109--126 

\bibitem[\protect\citeauthoryear{Madjarska et al.}{2011}]{madrajska2011} Madjarska, M.; Vanninathan, K.; Doyle, J. G.; 2011; Astronomy \& Astrophysics; 532; L1; 4pp

\bibitem[\protect\citeauthoryear{Mumfort et al.}{2013}]{mumfort2013} Mumford, S.; P\'{e}rez-Su\'{a}rez, D.; Christe, S.; et al.; 2013; Proceedings of the 12th Python in Science Conference; 74 – 77

\bibitem[\protect\citeauthoryear{Murawski et al.}{2011}]{murawski2011} Murawski, K.; Srivastava, A. K.; Zaqarashvili, T. V.; 2011; Astronomy \& Astrophysics; 535; A58; 9pp

\bibitem[\protect\citeauthoryear{Nistic\'{o} et al.}{2015}]{nistico2015} Nistic\'{o}, G.; Zimbardo, G.; Patsourakos, S.; et al.; 2015; Astronomy \& Astrophysics; 583:10--20

\bibitem[\protect\citeauthoryear{Parenti et al.}{2002}]{parenti2002} Parenti, S.; Bromage, B. J. I.; Bromage, G. E.; 2002; Astronomy \& Astrophysics; 384:303--316

\bibitem[\protect\citeauthoryear{Penza et al.}{2006}]{penza2006} 	
Penza, V.; Pietropaolo, E.; Livingston, W.; 2006; Astronomy \& Astrophysics; 454:349--358 

\bibitem[\protect\citeauthoryear{Pesnell et al.}{2012}]{pesnell2012} 	
Pesnell, W. D.; Thompson, B. J.; Chamberlin, P. C.; 2012; Solar Physics; 275:3-15 

\bibitem[\protect\citeauthoryear{Pike \& Harrison}{1997}]{pike1997}  Pike, C. D.; Harrison, R. A.; 1997; Solar Physics; 175:457--465

\bibitem[\protect\citeauthoryear{Pike \& Mason}{1998}]{pike1998} Pike, C. D.; Mason, H. E.; 1998; Solar Physics; 182:333

\bibitem[\protect\citeauthoryear{Scullion et al.}{2010}]{scullion2010} Scullion, E.; Doyle, J. G.; and Erd\'{e}lyi, R.; 2010; Memorie della Societa Astronomica Italiana; 81:737

\bibitem[\protect\citeauthoryear{Secchi}{1877}]{secchi1877} Secchi, P. A.; 1877; Le Soleil; Vol. 2 (Paris: Gauthier-villas) 

\bibitem[\protect\citeauthoryear{Sekse et al.}{2012}]{sekse2012} Sekse, D. H.; Rouppe, L. H. M.; de Pontieu, B.; 2012; The Astrophysical Journal; 752:108--122

\bibitem[\protect\citeauthoryear{Shetye et al.}{2015}]{shetye2015} Shetye, J.; Tripathi, D.; Dikpati, M.; 2015; The Astrophysical Journal; 799:220--231

\bibitem[\protect\citeauthoryear{Sterling}{2000}]{sterling2000} Sterling, A. C.; 2000; Solar Physics; 196:79--111

\bibitem[\protect\citeauthoryear{Sterling et al.}{2010}]{sterling2010} Sterling, A. C.; Harra, L. K.; Moore, R. L.; 2010; The Astrophysical Journal; 722:1644--1653

\bibitem[\protect\citeauthoryear{Sterling et al.}{2015}]{sterling2015} Sterling, A. C.; Moore, R. L.; Falconer, D. A.; et al.; 2015; Nature; 523:437--470

\bibitem[\protect\citeauthoryear{Sterling et al.}{2016}]{sterling2016} Sterling, A. C.; Moore, R. L.; Falconer, D. A.; et al.; 2016; The Astrophysical Journal; 821:100--117

\bibitem[\protect\citeauthoryear{Vernazza et al.}{1981}]{vernazza1981} Vernazza, J. E.; Avrett, E. H.; Loeser, R.; 1981; Astrophysical Journal Supplement Series; 45:635--725

\bibitem[\protect\citeauthoryear{Zaqarashvili et al.}{2010}]{zaqarashvili2010} Zaqarashvili, T. V.; Carbonell, M.; Oliver, R.; Ballester, J. L.; 2010; The Astrophysical Journal; 724:95--98

\bibitem[\protect\citeauthoryear{Zaqarashvili \& Erd\'{e}lyi}{2009}]{zaqarashvili2009} Zaqarashvili, T. V.; Erd\'{e}lyi, R.; 2009; Space Science Reviews; 149:355--388
\end{thebibliography}
\end{document}